\documentclass[useAMS,usenatbib,rotating]{mnras}
\usepackage{amssymb}     \usepackage{amsmath}    \usepackage{amsfonts}
\usepackage{xspace}    \usepackage{longtable}    \usepackage{graphicx}
\usepackage{times}

\newcommand{\HI}{\mbox{\sc H{i}}}

\newcommand{\Htwo}{\mbox{\sc H$_2$}}

\newcommand{\kms}{\mbox{km s$^{-1}$}}
\newcommand{\msun}{\mbox{$M_\odot$}}

\newcommand{\vmax}{\mbox{$V_{\rm{max}}$}}
\newcommand{\Qsg}{\mbox{$Q_{\rm{sg}}$}}
 
\title[Characterising SFE with marginally-stable disks]
  {Characterising uniform star formation efficiencies with marginally-stable galactic disks}
\author[Wong, O.I. et al.]
  {O.\ Ivy~Wong,$^{1,2}$ G.R.~Meurer,$^{1,3}$ Z.~Zheng,$^{4}$   T.M.~Heckman,$^{5}$ D.A.~Thilker$^{5}$ 
\newauthor \& M.A.~Zwaan$^{6}$ \\
  $^1$International Centre for Radio Astronomy Research, ICRAR M468, 35 Stirling Highway, Crawley, WA 6009, Australia\\
  $^2$ARC Centre of Excellence for All-Sky Astrophysics (CAASTRO)\\
  $^3$School of Physics, University of Western Australia, 35 Stirling Highway, Crawley, WA 6009, Australia\\
  $^4$National Astronomical Observatories, Chinese Academy of Sciences, 20A Datun Rd, Beijing 100012, China\\
  $^5$Department of Physics and Astronomy, Johns Hopkins University, 3701 San Martin Drive, Baltimore, MD 21218, USA\\
  $^6$European Southern Observatory, Karl-Schwarzschild-Strasse 2, D-85748 Garching bei M\''unchen, Germany\\
}
\date{Released 2016 Xxxxx XX}

\pagerange{\pageref{firstpage}--\pageref{lastpage}} \pubyear{2016}

\def\LaTeX{L\kern-.36em\raise.3ex\hbox{a}\kern-.15em
    T\kern-.1667em\lower.7ex\hbox{E}\kern-.125emX}

\begin{document}

\label{firstpage}

\maketitle

\begin{abstract} 
We examine the \HI-based star formation efficiency ($SFE_{\HI}$), the ratio of star formation rate to the atomic Hydrogen (\HI) mass, in the context of a constant stability star-forming disk model.  Our observations of \HI-selected galaxies show $SFE_{\HI}$ to be fairly constant (log~$SFE_{\HI}=-9.65$~yr$^{-1}$ with a dispersion of 0.3 dex) across $\sim 5$ orders of magnitude in stellar masses.  We present a model to account for this result, whose main principle is that the gas within galaxies forms a uniform stability disk and that stars form within the molecular gas in this disk.  We test two versions of the model differing in the prescription that determines the molecular gas fraction, based on either the hydrostatic pressure, or the stellar surface density of the disk. For high-mass galaxies such as the Milky Way, we find that either prescription predicts $SFE_{\HI}$ similar to the observations.  However, the hydrostatic pressure prescription is a more accurate $SFE_{\HI}$ predictor for low-mass galaxies. Our model is the first model that links the uniform $SFE_{\HI}$ observed in galaxies at low redshifts to star-forming disks with constant marginal stability.  While the rotational amplitude \vmax\ is the primary driver of disk structure in our model, we find the specific angular momentum of the galaxy may play a role in explaining a weak correlation between $SFE_{\HI}$ and effective surface brightness of the disk.

\end{abstract}


\section{Introduction} 

While star formation is largely understood to be a process local to within individual 
galaxies, the strong correlation between the observed integrated star formation rates (SFR) and 
total stellar masses ($M_{*}$) at both low- and high-redshifts have led astronomers to assume that 
this is due to a fundamental universal star formation efficiency (SFE), 
defined as the star formation rate normalised by the gas mass \citep[e.g.\ ][]{brinchmann04,salim07,schiminovich10,karim11,whitaker12,sobral14,hunt15,popping15}.

Sub-kiloparsec scale studies of nearby galaxies have found the $SFE$ to be constant when the $SFE$ is measured 
in terms of the molecular \Htwo\ gas mass \citep[e.g.\ ][]{leroy08}.  While this is true for 
the \Htwo-normalised $SFE_{\rm{H2}}$, a similar study by \citet{bigiel08} found the relationship between
the total gas density (sum of both molecular and atomic gas) and the SFR density to show
large variation within and between different spiral galaxies.  \citet{bigiel08} attribute this
variation to the variation of the \Htwo/\HI\ ratio ($R_{\rm{mol}}$) as a function of local environmental
factors. Recent observations of massive galaxies (stellar mass, $M_* > 10^{10}$ M$_{\odot}$) indicate a near 
constant integrated \HI-based $SFE$ ($SFE_{\HI}$) averaging 10$^{-9.5}$ year$^{-1}$\citep{schiminovich10,hunt15}. 
Similarly, \citet{huang12} found a weakly positive correlation between the $SFE_{\HI}$ and the 
stellar mass of galaxies for an \HI-selected sample of galaxies (which typically consists of 
galaxies less massive than $10^{10}$ M$_{\odot}$). This is rather surprising since the \Htwo/\HI\ 
varies strongly within galaxies and from galaxy to galaxy 
\citep{tacconi86,wong02,blitz06,bigiel08,leroy08}.  
{\em{Why is the observed global $SFE_{\HI}$ so uniform across all star-forming galaxies?}}


In this paper, we construct a model of a galactic disk with a uniform disk
stability \citep{toomre64,zheng13} and compare the predicted $SFE_{\HI}$ to those observed in nearby galaxies from the 
the Survey for Ionization in Neutral Gas Galaxies \citep[SINGG; ][]{meurer06}
 and the Survey of Ultraviolet emission in Neutral Gas Galaxies \citep[SUNGG; ][]{wong07}.
We find that the observed uniformity in  global $SFE_{\HI}$ across 5 magnitudes of stellar masses can be 
reproduced by our constant-$Q$ disk model.
This is the third paper in a series in which we develop the uniform disk stability model to
 first account for the gas distribution beyond the optical disk \citep{meurer13}, and then the
 gas and star formation distribution within the optical disk \citep{zheng13}. Here we model an
 integrated property, $SFE_{\HI}$, that connects the inner and outer disk and use a much larger 
comparison sample than our previous studies.

Section 2 describes the methods used to construct our stable galactic disk models.  
 Section 3 presents the 
sample of nearby star-forming galaxies used as a comparison to our
model star-forming disks.  Section 4 discusses the results of our models and compares the  model
results to those measured from observations of nearby star-forming galaxies. We present 
our conclusions in Section 5.


\section{Modelling the star formation efficiency with constant Q disks}
Here we explore whether the observed behaviour of $SFE_{\HI}$  can be explained by a model based on the hypothesis that galaxy disks evolve toward a state of constant disk stability parameter $Q$ \citep{toomre64}.  Recently, \citet{meurer13} showed that hypothesis can explain how \HI\ traces dark matter in the outer parts of galaxies \citep[the so called ``Bosma relation''; ][]{bosma81} by using a single-fluid $Q$ model.  
On the other hand, \citet{zheng13} showed that by assuming a two-fluid (stars and gas) version of the stability parameter $\Qsg$ is constant and adopting reasonable assumptions about the Star Formation Law (SFL) we can account for the distribution of the different gas phases and the star formation in  
nearby galaxies. Both \citet{meurer13} and \citet{zheng13} dealt with the radial
properties of small representative samples of disk galaxies.  Here we adapt these models
 and apply them to a much larger sample of galaxies in order to understand integrated
properties, in particular the star formation efficiency, along the full range of 
star forming galaxies.

\subsection{Model overview}

In this paper, we use a modified version of the constant two-fluid \Qsg\ model
\citep{zheng13} and compare these model results to a different low-redshift sample
of galaxies which samples a larger range and variety of star-forming galaxies.
Following the results  of \citet{zheng13}, we assume that the gas and stars are both
in cold disks; and adopt a constant \Qsg=1.6 and fix the gas velocity dispersion  
($\sigma_g$) at 11~\kms, following the results of the THINGS survey where the gas velocity dispersion is derived from the 2nd order \HI\ moment maps \citep{leroy08,zheng13}.  
We adopt a Universal Rotation Curve \citep{persic96,battaner00} to specify the 
galaxy dynamics and adopt observationally-based scaling relations to specify the 
stellar mass distribution.  We are then in a position to calculate the cold gas 
distribution, and after separating the gas into molecular and atomic phases,  we
then calculate the SFR distribution.   Here, we consider only the most  
successful two prescriptions from \citet{zheng13} for aportioning the gas into 
the atomic and molecular phases.

\begin{figure}
\begin{center}
\includegraphics[scale=.4]{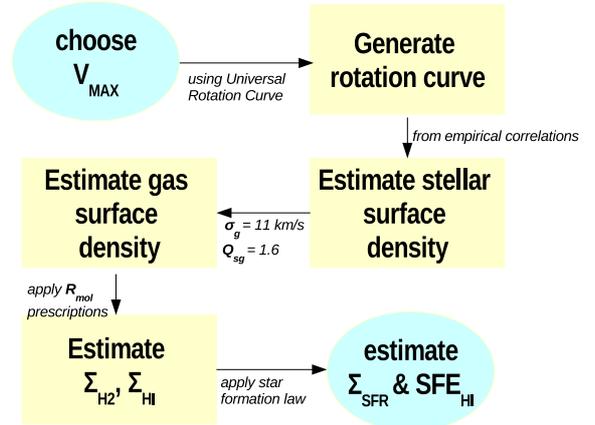}
\end{center}
\caption{{\bf{Flow diagram summarising our method for determining $SFE_{\HI}$ in our model. }}}
\label{flowchart}
\end{figure}

To be more specific, each model galaxy is parameterised as a function of the 
rotation curve amplitude, \vmax.  
As such, we describe the properties of our model galaxies as one-dimensional radial profiles.  
The stellar mass distribution in our models are determined via two empirically observed
relationships from our sample, namely the Tully-Fisher relationship and the relationship 
between the stellar surface brightness and \vmax.  We then solve for the gas 
distribution ($\Sigma_g$) using the two-fluid $Q$ approximation described by \citet{wang94}.
  From the resulting $\Sigma_g$, we apportion the gas into its molecular and atomic 
components using two different prescriptions for the molecular to atomic gas ratio ($R_{\rm{mol}}$)
via: 1) the stellar surface density ; and 2) the hydrostatic pressure.  Subsequently, we
estimate the SFR from the molecular gas content by assuming a largely linear relationship 
between \Htwo\ and the SFR.
Figure~\ref{flowchart} shows a flow diagram that summarises how our model determines the 
 $SFE_{\HI}$ from our model.  More details of our model is described in Section 2.2.

In summary our model is essentially a prescription for how to distribute the gas phases
and star formation within galaxies.  The essential principle of the model is that galaxy
disks evolve to have a uniform stability parameter.  In order to calculate the model, the
rotation curve of the galaxy and distribution of existing stars is required.  There are
many important details that this model does not address, such as the mechanism by which
 the stability parameter is maintained.  Presumably it involves feedback between the 
energy input from supernovae and stellar winds from recent star formation.  Recent models
 by \citet{lehnert13} argue that star formation-driven turbulence drives the observed 
velocity dispersions and that the disk instability is a result of such turbulence.
However, how the gas reservoir responds to maintain $Q$, and the exact value of $Q$ that 
results are not explicitly addressed.  Similarly we use known scaling relations to 
determine the stellar distribution.  But we do not address the underlying physics of these
 relations.  In section 5.1, we show that the model results remain within the scatter of
 the observations when adjusting the relevant free parameters of the model within 
known constraints.

\subsection{Model details}

We employ the \citet{wang94} approximation to specify the two fluid stability parameter \Qsg:
\begin{equation}
\frac{1}{\Qsg} = \frac{1}{Q_s} + \frac{1}{Q_g}
\end{equation}
where $Q_s$ and $Q_g$ are the Toomre stability parameters considering only the stars and gas respectively. For the stars we have 
\begin{equation}
Q_s = \frac{\kappa\sigma_{\star,r}}{\pi\Sigma_s},
\end{equation}
where $\kappa$ is the epicyclic frequency, $\sigma_{\star,r}$ is the velocity dispersion of the stars in the radial direction (in a cylindrical coordinate system), and $\Sigma_s$ represents the projected surface mass density of stars. Similarly for the gas we have
\begin{equation}
Q_g = \frac{\kappa\sigma_g}{\pi\Sigma_g},
\end{equation}
where $\sigma_g$ is the one-dimensional velocity dispersion of the gas and $\Sigma_g$ the surface mass density of gas. 


We model the $SFE_{\HI}$ relation as a one-dimensional sequence as a function of rotation curve amplitude \vmax . This is used as the driving parameter since it typically is used as short-hand for the halo mass in a standard ($\Lambda$) Cold Dark Matter cosmology \citep[e.g.\ ][]{navarro97}. We specify the rotation curve shape $V_c(R)$ using the ``Universal Rotation Curve'' (URC) algorithm of \citet{persic96}, as implemented by \citet{battaner00}. This family of rotation curves is based on long-slit observations of disk galaxies which range in $v_{\rm{max}}$ from 60--280~km~s$^{-1}$ \citep{mathewson92}.  The stellar distribution in each galaxy is assumed to be a pure exponential disk \citep[e.g.\ ][]{freeman70}; the total stellar luminosity for given \vmax\ is given by the Tully-Fisher relationship \citep[TFR; ][]{fisher81,meyer08} while the central surface brightness, and thus scale length are given by the surface brightness -- \vmax\ relationship, which is a variant of the well-known surface brightness -- luminosity relationship \citep[e.g.\ ][]{disney85,kauffmann03}. These two relationships are defined in the $R$-band using SINGG data, and so are well-calibrated to our sample. We note that our model does not take into consideration the contribution from a bulge component. Further discussion on the implications of our ``bulge-less'' model can be found in Section 5.

In order to convert luminosity to stellar mass we multiply by the mass to light ratio $M_*/L_R$, noting that $M_R=4.61$\footnote{All magnitudes are expressed in the ABmag system in this work.} is the absolute magnitude of the Sun in the $R$-band. Star-forming and passively evolving galaxies are easily distinguished in the optical colour-magnitude space forming the `blue cloud' and the `red sequence' \citep[e.g.\ ][]{blanton03,bell03,driver06}.   There is a distinct tilt to the `blue cloud' in most optical and ultraviolet colours with the most luminous galaxies being noticeably redder than the dwarfs. Almost all SINGG and SUNGG galaxies reside in the `blue cloud'. However we do not have optical photometry for most galaxies except in the $R$-band. So we use typical SUNGG galaxies with published optical $B-R$ photometry from the NASA/IPAC Extragalactic Database (NED)\footnote{The NASA/IPAC Extragalactic Database (NED) is operated by the Jet Propulsion Laboratory, California Institute of Technology, under contract with the National Aeronautics and Space Administration.} at the luminosity extremes of the blue sequence to derive our adopted relationship between $M_*/L_R$ and absolute magnitude $M_R$
\begin{equation}
log(M_*/L_R) = -1.578 - 0.0856 M_R \label{e:mlr}
\end{equation}
where $M_R$ is in AB magnitude and the resultant $M_*/L_R$ is in solar units. This relationship was derived from the \citet{bell03} stellar population models which assume a ``diet Salpeter'' IMF. 

\begin{figure}
\begin{center}
\includegraphics[scale=.43]{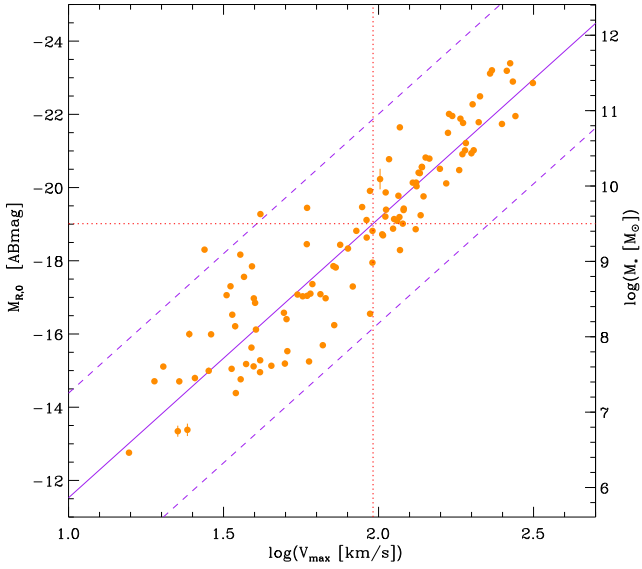}
\end{center}
\caption{The Tully-Fisher Relationship for SINGG galaxies in the $R$ band, showing the dust corrected absolute $R$-band magnitude $M_{R,0}$ derived from the SINGG images versus the rotation velocity amplitude $V_{\rm max}$ derived from \HI\ line widths from the HIPASS survey. Selection criteria for inclusion in this plot and references to the corrections employed are given in the text. An ordinary least squares fit to the data, with iterative clipping of outliers is shown as the solid line. The dashed lines show the $\pm 3\sigma_R$ clipping limit to the fit, where $\sigma_R$ is the dispersion about the fit. The right axis converts $M_{R,0}$ to stellar mass using eq.~\ref{e:mlr}.  The horizontal dotted line shows the stellar mass below which \citet{simons15} find the TFR breaks down.  The vertical dotted line marks where this intercepts our adopted TFR. }
\label{f:singg_tfr}
\end{figure}

Figure~\ref{f:singg_tfr} shows the SINGG based TFR. The sample of galaxies used is single galaxies that have axial ratios $1.6 < a/b < 6$. Total luminosities were measured in an aperture large enough to contain all the $R$-band light in the SINGG images. Luminosities are corrected for foreground and internal dust absorption using the algorithms in \citet{meurer06}, which also compiles the adopted distances. The \vmax\ values are derived from \HI\ line widths at 50\%\ peak intensity from the HIPASS survey \citep{meyer04,koribalski04} as compiled by \citet{meurer06}. We follow \citet{meyer08} in correcting the line widths for relativity, instrumental resolution, turbulence, and inclination which we derive following the prescription of \citet{meurer06}. We make an ordinary least squares fit of absolute R band magnitude $M_R$ as a function of $\log(\vmax)$ to yield our adopted TFR:
\begin{equation}
M_R = -3.90 - 7.622\log(\vmax). \label{e:tfr}
\end{equation}
The adopted fit is shown as the solid line in Figure~\ref{f:singg_tfr}, while the dashed lines are offset by $\pm 3\sigma$ where $\sigma = 0.95$ mag is the scatter about the fit. Points outside of{\bf{ $\pm 3\sigma$}} from the fit are iteratively clipped in our fitting procedure. 

This paper is not focussed on the TFR; the measurements are not ``tuned'' to provide the tightest most accurate relationship.  Nevertheless, the TFR we find is of reasonable quality and reaches well in to the dwarf galaxy regime. Several modern studies have shown a break in the TFR for low mass galaxies. \citet{mcgaugh00} and \citet{mcgaugh05}  note that the TFR changes slope and has larger scatter for stellar mass $M_\star \lesssim 10^9\, \msun$. They find a tighter more continuous relationship between the baryonic mass (stellar plus gas mass) and the rotational amplitude which they coined the baryonic Tully Fisher Relationship. Recently \citet{simons15} and \citet{cortese14} found that for $M_\star \lesssim 10^{9.5}\, \msun$ there are more extreme outliers to the TFR than at higher masses; and that using the kinematic quantity $S_{0.5}$ which combines contributions from rotation and turbulence restores a tighter relationship. We find an increased scatter about our TFR for dwarf galaxies. For $\vmax \le 96\, {\rm km\, s^{-1}}$ (corresponding to $M_R > -19.02$ ABmag, and $M_\star < 10^{9.5}\, \msun$ following eqs. \ref{e:tfr} and \ref{e:mlr}) the residuals have a dispersion of 1.25 mag, while the dispersion for the points with $\vmax > 96\, {\rm km\, s^{-1}}$ is 0.72 mag (no clipping was done). There is no perceptible change in slope in the TFR for dwarfs within the scatter of our measurements. \citet{meyer08}  attribute the increased scatter in the TFR of HIPASS galaxies (our parent sample) to the difficulty of estimating the inclination of dwarfs. We likely avoid the extreme outliers found by \citet{simons15}  because we are using an \HI\ selected sample, pruned of strongly interacting galaxies, with kinematics based on \HI\ line widths. Like their $S_{0.5}$ parameter these widths are determined by contributions from rotation and turbulence; additionally they sample the dynamics to larger radii where the rotation curve is flatter than the ionized gas tracers used by \citet{simons15} and \citet{cortese14}.  We conclude that our adopted TFR is adequate for our purposes. 

\begin{figure}
\begin{center}
\includegraphics[scale=.43]{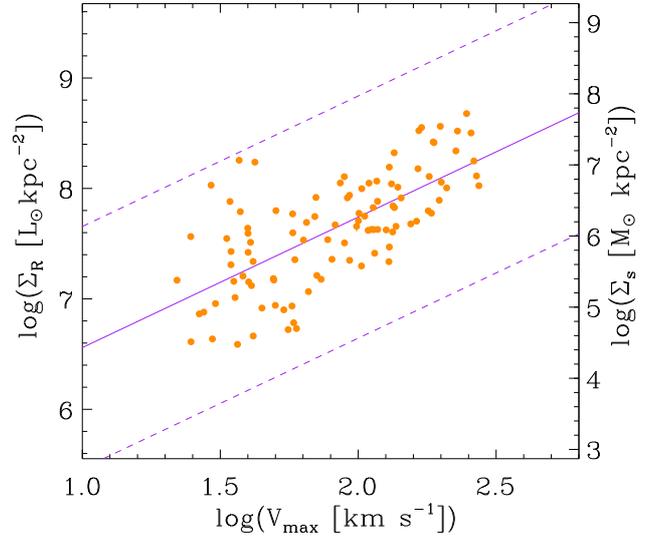}
\end{center}
\caption{The observed relationship between $R$-band effective surface brightness $\Sigma_R$ and rotation velocity amplitude for SINGG galaxies.  An ordinary least squares fit to the data, is shown as the solid line. The dashed lines are offset by $\pm 2\sigma_\Sigma$ from the fit, where $\sigma_\Sigma$ is the dispersion about the fit. The right axis shows the approximate stellar mass density, $\Sigma_s$.  }
\label{f:vrot_sb}
\end{figure}

Figure~\ref{f:vrot_sb} shows the $R$-band $\Sigma_R$ versus \vmax\ relationship for SINGG galaxies.
Here $\Sigma_R$ is the effective surface brightness - that is the average face-on surface brightness within an aperture containing half of the $R$-band luminosity. Employing an ordinary least squares fit to these data yields
\begin{equation}
log(\Sigma_R) = 8.14 + 1.176\log(\vmax).\label{e:sbrv}
\end{equation}
The adopted fit is shown as the solid line, while the dashed lines are offset by $\pm 2\sigma$ where $\sigma = 0.37$ dex is the scatter about the fit. There are no points outside of $\pm 2\sigma$ from the fit. The right axis gives the approximate scaling to stellar mass density $\Sigma_s$, as derived by combining eqs.\ \ref{e:mlr}, \ref{e:tfr} and \ref{e:sbrv}. With the assumption that to first order the surface brightness distribution is a single exponential disk \citep{freeman70} and the standard relationships between effective surface brightness and extrapolated central surface brightness, then the disk scale length is given by
\begin{equation}
R_d = \sqrt{\frac{10^{-0.4(M_R - 4.61)}}{5.647\pi\Sigma_r}}.
\end{equation}
We follow the prescription given in equation B3 of \citet{leroy08} to specify the stellar velocity dispersion in the $z$ direction, $\sigma_{\star,z}$ from $R_d$ and the local stellar surface mass density $\Sigma_s$ which is obtained from $\Sigma_R$ after applying eq.~\ref{e:mlr}.


As noted in \citet{meurer13}, a constant $Q$ disk dominated by gas in
the outskirts of galaxies where the rotation curve is flat should have
$\Sigma_g(R) \propto R^{-1}$. The corresponding integrated gas content
is not finite unless the distribution is truncated. Here we truncate the
disks at the radius where the orbital time $t_{\rm orb} = 2\pi R/V_c(R) = 1$ Gyr, following 
the work of \citet{meurer15} on the relationship
between $V_c$ and maximum radius in galaxies including the SINGG and
SUNGG sample.
 
The total gas radial profile $\Sigma_g$ is apportioned to molecular and atomic components using models for the molecular to atomic ratio 
\begin{equation}
R_{\rm mol}\equiv \Sigma_{\rm H_2}/\Sigma_{\rm HI}
\end{equation}
Following \citet{zheng13}, we test two different prescriptions for $R_{\rm mol}$, both of which were also trialled by \citet{leroy08}. The first is a purely empirical scaling of the stellar surface mass density ($\Sigma_\star$) where 
\begin{equation}
R_{\rm mol,s}=\frac{\Sigma_\star}{81\, M_{\odot}\, pc^{-2}}.
\end{equation}
The other depends upon the  hydrostatic pressure, $P_h$, and is
\begin{equation}
R_{\rm mol,p}= \left(\frac{P_h}{1.7\times10^4\, cm^{-3}\,K\, k_B} \right)^{0.8},
\end{equation}
where $k_B$ is the Boltzmann constant and $P_h$ is given by \citet{elmegreen89} as
\begin{equation}
P_h= \frac{\pi}{2}G \Sigma_g(\Sigma_g+\frac{\sigma_g}{\sigma_{\star,z}}\Sigma_\star),
\end{equation}
where $\sigma_{\star,z}$ is the vertical stellar velocity dispersion.

\begin{figure}
\begin{center}
\includegraphics[scale=.63]{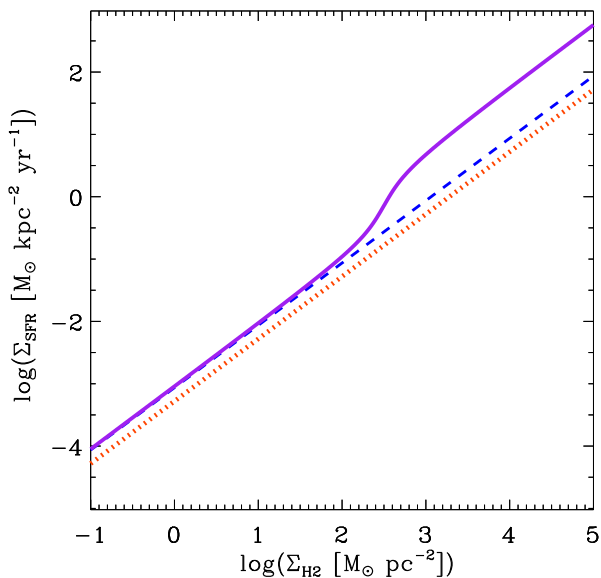}
\end{center}
\caption{Our adopted molecular Star Formation Law, eq.~\ref{e:sfl} is shown as the purple (solid) line. The linear molecular SFLs of \citet{bigiel08} and \citet{zheng13} are shown by the dotted orange and dashed blue lines, respectively.}
\label{f:sfls}
\end{figure}

The radial profile of star formation $\Sigma_{\rm SFR}$ is calculated using an SFL that depends only on the molecular gas content $\Sigma_{\rm H2}$. \citet{bigiel08} used data on normal spiral and dwarf galaxies from the THINGS survey to show that in the bright part of galaxies a linear molecular SFL holds: $\Sigma_{\rm SFR} \propto \Sigma_{\rm H2}$. They find the ratio between the two $\Sigma_{\rm SFR}/\Sigma_{\rm H2} = {\rm SFE_{H2}} = 5.25\times 10^{-10}\, {\rm yr^{-1}}$. \citet{zheng13} used a subset of the same data and a different algorithm to fit a uniform linear molecular SFL and find ${\rm SFE_{H2}} = 8.66\times 10^{-10}\, {\rm yr^{-1}}$.
\citet{bigiel08} show that their linear molecular SFL is well calibrated only over 1.5 dex in gas density: $log(\Sigma_{\rm H2}/(1\, \msun\, {\rm pc^{-2}})) = 0.5$ to 2.0. While this range covers the range of gas densities typically seen in normal star forming galaxies, they show in their Fig.\ 15 that the linear molecular SFL underpredicts $\Sigma_{\rm SFR}$ in starburst galaxies \citep[as observed by][]{kennicutt98} which have ${\rm SFE_{H2}} \sim 6.0 \times 10^{-9}\, {\rm yr^{-1}}$ and typically significantly higher $\Sigma_{\rm H2}$ compared to normal galaxies.  Therefore, we adopt an SFL that transitions between a low ${\rm SFE_{H2}}$, set at the value found by \citet{zheng13}, and a more efficient ${\rm SFE_{H2}}$ set at the value implied by the \citet{kennicutt98} starburst data.  The transition occurs at $log(\Sigma_{\rm H2}/(1\, \msun\, {\rm pc^{-2}})) = 2.5$ and we use the arctan function as a convenient form to transition between the cases. The exact functional form we employ is 
\begin{equation}
\Sigma_{\rm SFR} = 8.66\times 10^{-4} \Sigma_{\rm H2} 10^{0.84(\arctan(5(\log(\Sigma_{\rm H2})-2.5))/\pi + 0.5)}\label{e:sfl},
\end{equation}
Where $\Sigma_{\rm SFR}$ is in units of \msun\ kpc$^{-2}$ yr$^{-1}$ and $\Sigma_{\rm H2}$ is in units of \msun\ pc$^{-2}$ yr$^{-1}$, and includes contributions from helium and heavy elements. Fig.~\ref{f:sfls} plots this SFL in comparison to the linear molecular SFLs of \citet{bigiel08} and \citet{zheng13}. The asymptotic linear behavior of this SFL at normal and high densities yields molecular gas consumption times of 1.2 Gyr and 0.17 Gyr respectively.

\subsection{Model demonstration}
\label{s:demon}
\begin{figure*}
\begin{center}
\includegraphics[scale=.85]{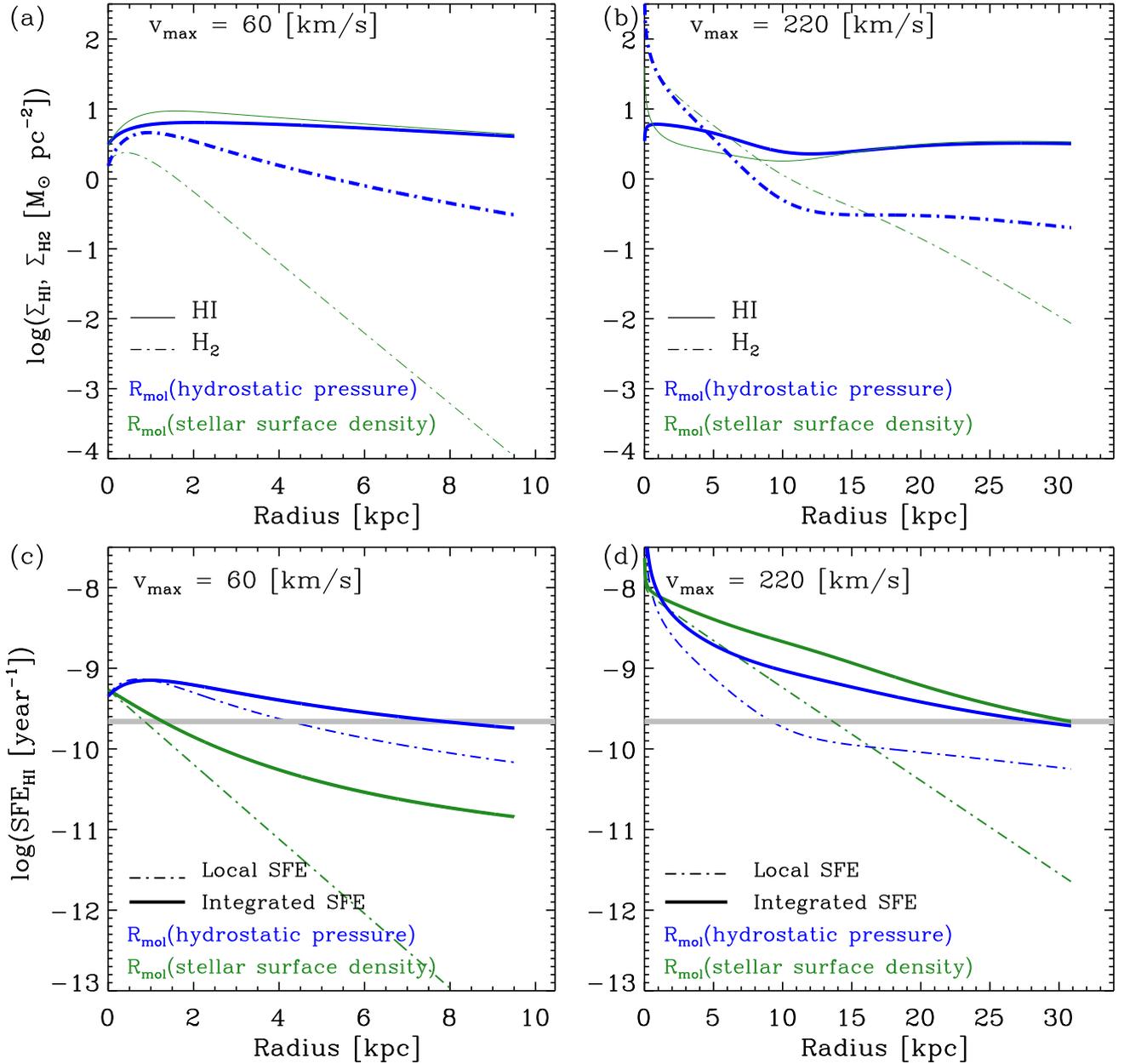}
\end{center}
\caption{Panels (a) and (b) show the gas surface mass density as a function 
of radius for two constant-$Q$ disk models: (a)
 a dwarf galaxy with a maximum circular velocity of 60 \kms; 
and (b) a massive disk galaxy with a maximum circular velocity of 220 \kms.  
The atomic \HI\ gas density ($\Sigma_{\rm{HI}}$) is represented by the solid 
lines and the molecular \Htwo\ gas density ($\Sigma_{\rm{H2}}$) is represented by the
dot-dash lines. The models using the molecular gas ratio ($R_{\rm{mol}}$) 
estimated via the stellar mass surface density are shown in green, and those using
$R_{\rm{mol}}$ estimated via hydrostatic pressure arguments are shown in blue. 
Panels (c) and (d) show the local $SFE_{\HI}$ (dot-dashed lines) and integrated $SFE_{\HI}$ 
within the specified radius (solid lines) for our models. To guide
the eye, we show the mean $SFE_{\HI}$ (log~($SFE_{\HI}$)$=-9.65$) found from our  
sample  as a thick grey horizontal line.  }
\label{egmodel}
\end{figure*}

To illustrate the behaviour of our constant-$Q$ disk models, we show 
in panels (a) and (b) of Figure~\ref{egmodel} the \HI\ and \Htwo\ gas densities 
($\Sigma_{\rm{HI}}$, $\Sigma_{\rm{H2}}$) of our models for a low- and a 
high-mass galaxy. The left panels (a and c) of Figure~\ref{egmodel} represent
the models for a low-mass disk galaxy with a maximum circular velocity ($v_{\rm{max}}$)  
of 60~\kms.  The right panels (b and d) show a Milky Way analogue galaxy
having \vmax\ ~$=220$~kms$^{-1}$, close to the highest \vmax\ in our model sequence.
The models for the stellar mass density and hydrostatic pressure
 prescriptions for $R_{\rm{mol}}$ (section 2.1) are shown with green and blue lines,
 respectively.

In general, it appears that the stellar surface density $R_{\rm{mol}}$ prescription
is unable to convert the \HI\ to \Htwo\ at large radii where the stellar
surface densities are low. This is especially obvious in the case 
of the low-mass disk galaxy and at large disk radii for both low- and high- mass
disk galaxies.  These results of star formation at low gas densities are consistent 
with the observations of galaxies with  extended UV (XUV) disks \citep{thilker07} where 
star formation in the outer low-density regions is more common than expected,  
based upon our understanding of star formation within the established stellar disk.

The bottom row panels of Figure~\ref{egmodel} (panels (c) and (d)) show the
resulting  $SFE_{\HI}$ at a given radius (local) and the global $SFE_{\HI}$ interior 
to a radius (integrated) for our models of the low- and high-mass disk galaxies. 
Our modelled local $SFE_{\HI}$ radial profiles are consistent with recent observations 
 of nearby galaxies whereby the radial profiles of $SFE$ flatten in the outer parts of galaxies 
at much lower values than those in the optical disks \citep[e.g.\ ][]{yim15}.
At large radii, the integrated  $SFE_{\HI}$ for both low- and high-mass 
disk models using the hydrostatic disk pressure
 molecular gas fraction prescription are very close to the mean $SFE_{\HI}$ found
in our sample (as shown by the grey line in panels (c) and (d)).
While both $R_{\rm{mol}}$  prescriptions result in very similar integrated
$SFE_{\HI}$ for the model of the high-mass disk galaxy, this is not the case for the
model of the low-mass disk where  the stellar surface density prescription
underpredicts the observed integrated $SFE_{\HI}$. 

Previous studies of star formation rates (and efficiencies) with stellar surface densities
are likely to be biased towards  galaxies and portions of galaxies with high molecular gas and 
stellar densities \citep[e.g.\ ][]{wong13,leroy08} due to  optical/UV and molecular gas 
observational limits.  As such, only a narrow range of 
stellar surface densities (which trace the hydrostatic disk pressure) is studied.

\section{Sample and observations}
\begin{figure*}
\begin{center}
\includegraphics[scale=.21]{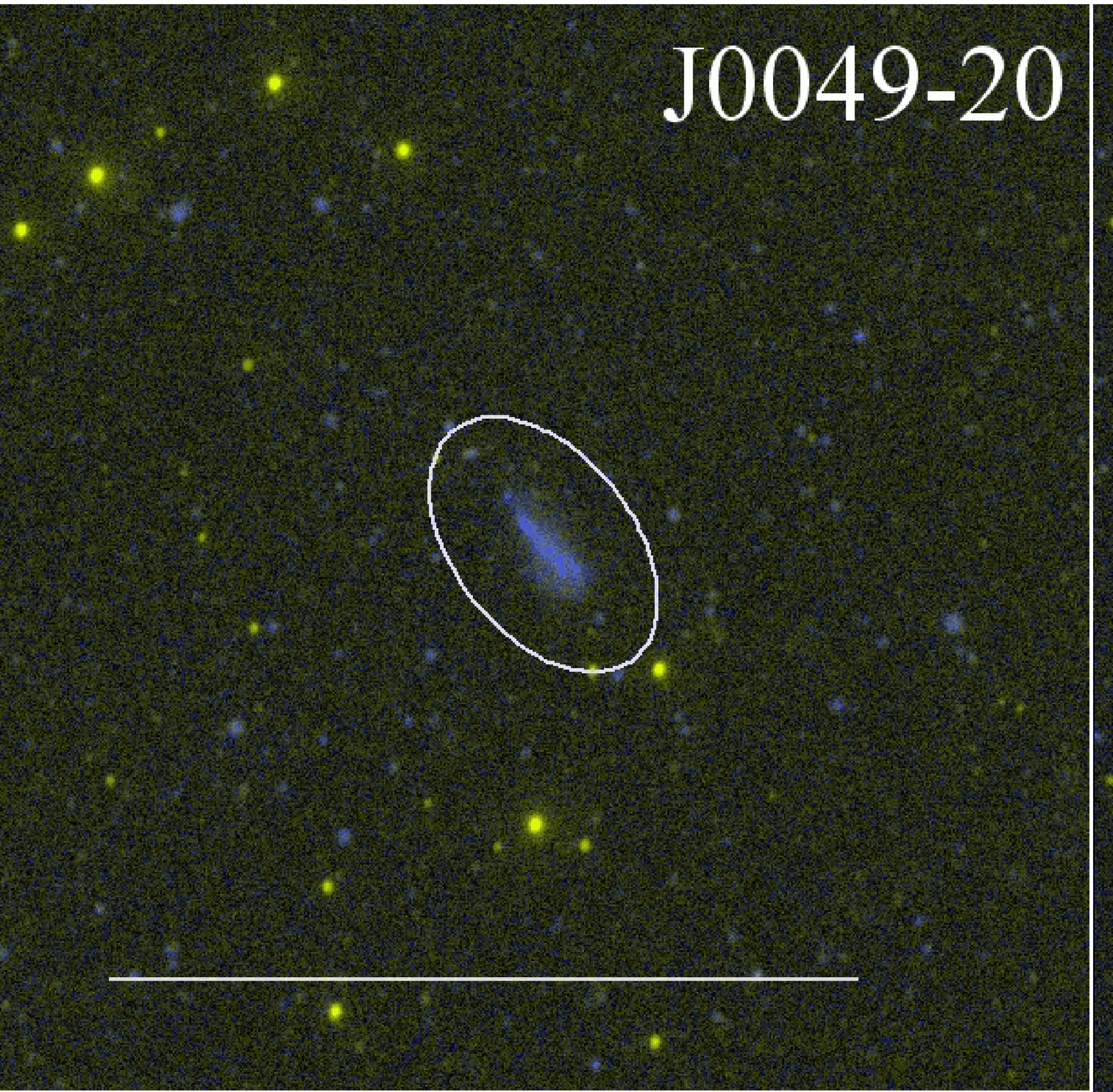}
\end{center}
\caption{Four galaxies representing the variety in size and stellar surface brightness
of the star-forming galaxies from the SUNGG survey.  GALEX three-colour 15 arcminute
by 15 arcminute postage stamps of the SUNGG UV observations where the FUV and NUV are 
represented by the blue and yellow, respectively. The RGB colour scaling of these
images are shown by  \citet{gildepaz07} where blue, green and red represents the FUV, 
linear combination of FUV\&NUV emission and NUV emission (as per \citet{gildepaz07}); 
respectively. The scale bar represents 10 kiloparsecs.
 }
\label{example}
\end{figure*}

We obtain our sample of nearby star-forming galaxies  from the Survey for Ionization
in Neutral Gas Galaxies \citep[SINGG; ][]{meurer06} and the Survey of Ultraviolet 
emission of Neutral Gas Galaxies \citep[SUNGG; ][]{wong07,wong15}---two surveys which 
imaged an \HI-selected sample of galaxies from the \HI\ Parkes All-Sky Survey
\citep[HIPASS; ][]{meyer04,zwaan04,koribalski04,wong06} in the optical and the ultraviolet 
wavelengths. The size of the sample used in this paper is a factor of seven greater than 
that used in \citet{meurer13} and \citet{zheng13} and spans a larger range of 
star-forming galaxies in size and brightness (see Figure~\ref{example}).
The optical $R$-band and H$\alpha$
 imaging for SINGG were primarily obtained from the 1.5-m telescope at the the Cerro
 Tololo Inter-American Observatory in Chile, while the Galaxy Evolution Explorer (GALEX)
 satellite telescope is used to obtain the far- and near-ultraviolet (FUV and NUV) for
 SUNGG at 1524-$\rm{\AA}$ and  2273 -$\rm{\AA}$, respectively.

The sample overlap between SINGG and SUNGG consists of 306 nearby galaxies.
We adopt as our sample the subset of those where (a) the 15~arcmin HIPASS beam 
contains only one associated star forming galaxy, and (b) the optical major to
minor axial ratio is $a/b > 1.6$. Multiple and face-on galaxies are thus excluded 
so that  $v_{\rm{max}}$ can be more accurately estimated from the \HI\ line width.

\begin{figure}
\begin{center}
\includegraphics[scale=.39]{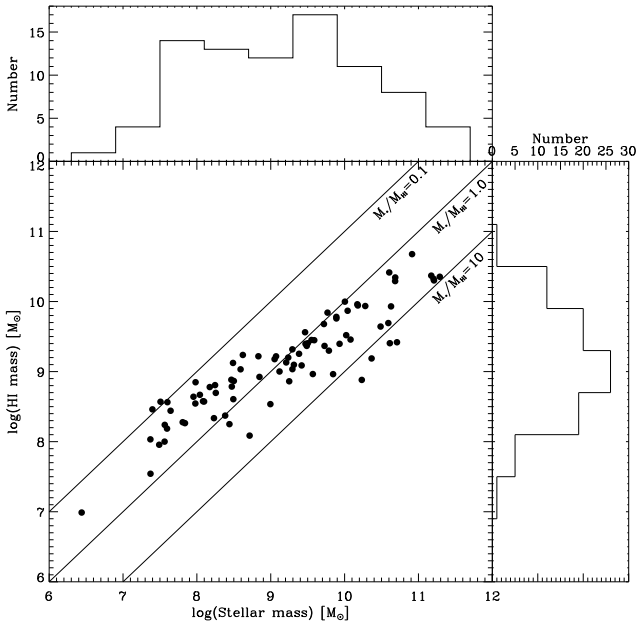}
\end{center}
\caption{The \HI\ and stellar mass distributions of our sample. The solid lines are lines of constant
stellar-to-\HI\ mass fractions ranging from 0.1 to 10.}
\label{massdist}
\end{figure}

Table~\ref{samp} lists the properties of the 84 galaxies used in this paper.
 The distances are from \citet{meurer06} and mostly calculated from their recessional
 velocities \citep[following ][]{mould00}.  They range from 3.3~Mpc to 112~Mpc with
 a median of 13~Mpc.
Figure~\ref{massdist} presents the \HI\ and stellar mass distributions
 for our \HI-selected sample of star-forming galaxies.  Distinct from optically-selected
 samples of star forming galaxies, we do not propagate a bias against star-forming galaxies 
with  low stellar masses and/or low stellar surface brightnesses.  
We note that approximately half of our sample have stellar masses below $10^9$~M$_{\odot}$,
 where there is an increased scatter in both the Tully-Fisher relationship and the 
 \vmax\ and $\Sigma_R$ relationship described in Section 2.2.


In this paper, we use the FUV emission from SUNGG as the tracer of the current star formation 
rate. The FUV emission is sensitive to the presence of $O$ and $B$ stars ($M_{*}>3$~M$_{\odot}$).
The other commonly-used star formation indicator is the $H\alpha$ emission which is sensitive to
$O$ stars ($M_{*}>20$~M$_{\odot}$). However, the ratio between FUV and H$\alpha$ can vary between 
galaxies  which may indicative of a varying initial mass function (IMF) or that the UV escape 
fraction may be different in low surface brightness and low mass galaxies \citep{meurer09,lee09}.
\citet{lee09} argue that the FUV emission is a more accurate tracer of the true star formation 
rate than the H$\alpha$ emission.  In this paper, we calculate the star formation rates from 
the FUV emission following 
\begin{equation}
SFR(FUV) = \frac{l_{\mathrm{FUV}}}{1.37 \times 10^{32} \mathrm{W\,\AA^{-1}}}
\end{equation}
which is eq.~2 of \citet{meurer09} adjusted for consistency with a
\citet{kroupa01} IMF.

The optical and UV 
magnitudes and luminosities have been corrected for both Galactic and internal 
dust extinction.  We have used the internal FUV attenuation correction method derived by 
\citet{salim07} for a sample of low-redshift galaxies. The internal dust absorption correction
for the optical observations are calculated empirically, based on the $R$-band luminosity 
\citep{meurer06,helmboldt04}.

\subsection{The effect of the internal dust corrections on the $SFE_{\HI}$}
Although the FUV emission is a well-established tracer of recent star formation, 
the FUV emission is highly susceptible to dust extinction and may 
artificially influence the observed range of $SFE_{\HI}$.  Figure~\ref{AInt}
compares the  $SFE_{\HI}$ as a function of circular velocity with 
and without internal dust extinction in Panels (a) and (b), respectively.
The standard deviation of the scatter in $SFE_{\HI}$ in panels (a) and (b)
are very similar with log values of 0.28 and 0.29, respectively. 

The omission of the dust extinction correction (as seen in Figure~\ref{AInt}b) results in 
$SFE_{\HI}$ which are on average 3.8 times lower than the average dust-corrected
 $SFE_{\HI}$ for our sample. This implies that on average, only $\sim26$\% of the FUV emission
escapes from the galaxies. 
While it is clear that the application of internal dust corrections have renormalised the 
$SFE_{\HI}$, the constancy and scatter of the $SFE_{\HI}$ is not
significantly reduced by these dust corrections. In Figure~\ref{AInt}
the colour of each data point shows the internal FUV dust correction
$A_{\mathrm FUV}$. $A_{\mathrm FUV}$ correlates with \vmax, we also
find a weak correlation of $A_{\mathrm FUV}$ with $SFE_{\HI}$.  These
are both expected - galaxies with more mass and more star formation
tend to be dustier.


 
\section{Star formation efficiency in nearby  galaxies}

\begin{figure}
\begin{center}
\includegraphics[scale=.4]{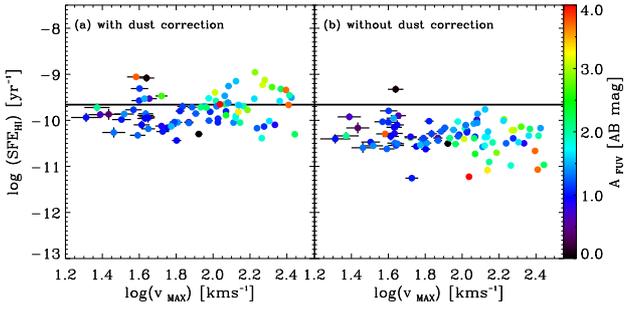}
\end{center}
\caption{Effect of internal dust extinction on the measured $SFE_{\HI}$ in our sample.
Panel (a) shows the integrated $SFE_{\HI}$ with internal dust correction
 as a function of $v_{\rm{MAX}}$ and panel (b) shows the integrated $SFE_{\HI}$ without
dust correction as a function of $v_{\rm{MAX}}$.  The  horizontal line
 marks the mean $SFE_{\HI}$ of our sample. The data points are coloured according to the
internal dust correction as shown on the colour bar on the right of the figure.}
\label{AInt}
\end{figure}

Figure~\ref{AInt}a presents the  star formation efficiencies of our sample galaxies as a function
 of  the circular velocity (\vmax).   Our sample have an mean and median log~$(SFE_{\HI})$ of $-9.65$ and $-9.81$, 
respectively, as well as a standard deviation of 0.28~dex. In comparison, the  stellar mass-selected 
GASS survey finds a constant
$SFE_{\HI}$ of  10$^{-9.5}$ year$^{-1}$ \citep{schiminovich10}, while the average log~$(SFE_{\HI})$ from ALFALFA
 is log~$(SFE_{\HI})=-9.95$ \citep{huang12}.

\clearpage
\onecolumn

\begin{center}
\begin{longtable}{lccccc}
\caption{Properties of our sample.}\label{samp}\\
\hline \hline \multicolumn{1}{l}{HIPASS galaxy}  &\multicolumn{1}{c}{Distance} & \multicolumn{1}{c}{$v_{\rm{MAX}}$} &  \multicolumn{1}{c}{$M_{\rm{R}}$} & \multicolumn{1}{c}{log $\Sigma_{\rm{R}}$} & \multicolumn{1}{c}{log($SFE_{HI}$)}   \\
\hline
\endfirsthead

\multicolumn{6}{c}%
{{\tablename \thetable{} -- continued from previous page}}\\
\hline \multicolumn{1}{l}{HIPASS galaxy}  &\multicolumn{1}{c}{Distance} & \multicolumn{1}{c}{$v_{\rm{MAX}}$} &  \multicolumn{1}{c}{$M_{\rm{R}}$} & \multicolumn{1}{c}{log $\Sigma_{\rm{R}}$} & \multicolumn{1}{c}{log($SFE_{HI}$)} \\
\hline
\endhead

\hline \multicolumn{6}{r}{Continued on next page}\\ \hline
\endfoot

\hline \hline
\endlastfoot

J0008-59             &  112.1 &   216. & -22.5 &  7.73 & -10.10 \\
J0014-23             &    7.0 &   106. & -19.2 &  7.33 &  -9.96 \\
J0039-14a            &   20.6 &    45. & -19.3 &  8.11 &  -9.53 \\
J0043-22             &    4.9 &    24. & -14.7 &  6.89 &  -9.72 \\
J0047-09             &   19.1 &    74. & -17.8 &  7.13 & -10.06 \\
J0047-25             &    3.9 &   209. & -21.8 &  8.02 &  -9.28 \\
J0049-20             &    3.3 &    21. & -12.8 &  6.72 &  -9.94 \\
J0052-31             &   22.9 &   173. & -22.0 &  8.31 & -10.17 \\
J0135-41             &    4.1 &    40. & -17.6 &  7.71 &  -9.57 \\
J0145-43             &    4.4 &    38. & -16.2 &  6.96 &  -9.06 \\
J0256-54             &    7.1 &    63. & -17.4 &  6.65 & -10.44 \\
J0307-31             &   66.9 &   189. & -21.8 &  7.54 & -10.13 \\
J0309-41             &   12.9 &    77. & -18.4 &  7.08 &  -9.72 \\
J0310-33             &   15.3 &    61. & -17.0 &  6.86 & -10.08 \\
J0317-37             &   13.5 &    27. & -14.7 &  7.53 &  -9.87 \\
J0320-52             &    7.0 &    43. & -17.0 &  7.52 &  -9.53 \\
J0328-08             &   17.4 &   129. & -20.1 &  7.47 & -10.05 \\
J0331-51             &   14.3 &   133. & -20.0 &  8.04 &  -9.15 \\
J0333-50             &    8.1 &    66. & -17.1 &  7.46 &  -9.80 \\
J0344-44             &   15.9 &   202. & -21.0 &  7.86 & -10.02 \\
J0345-35             &   10.8 &    39. & -14.8 &  6.55 & -10.32 \\
J0349-48             &   13.1 &   133. & -18.9 &  7.22 &  -9.90 \\
J0354-43             &   12.3 &    53. & -15.2 &  7.13 &  -9.47 \\
J0355-40             &   10.5 &    44. & -15.0 &  7.30 &  -9.95 \\
J0355-42             &   11.2 &   120. & -19.0 &  7.95 &  -9.62 \\
J0404-54             &   15.9 &   186. & -20.9 &  8.05 &  -9.23 \\
J0406-21             &   12.8 &    90. & -19.5 &  7.92 &  -9.70 \\
J0411-35             &   11.4 &    43. & -15.6 &  7.37 &  -9.89 \\
J0421-21             &   12.4 &    96. & -18.8 &  7.24 &  -9.81 \\
J0429-27             &   13.0 &    44. & -16.9 &  7.44 &  -9.08 \\
J0451-33             &   16.2 &   104. & -18.7 &  7.60 &  -9.58 \\
J0459-26             &   10.0 &   118. & -19.2 &  7.51 &  -9.81 \\
J0505-37             &   16.7 &   169. & -22.0 &  8.28 &  -8.96 \\
J0506-27             &   17.8 &    43. & -17.8 &  7.55 &  -9.96 \\
J0506-31             &   10.9 &    40. & -18.2 &  8.16 &  -9.31 \\
J0510-36             &   14.1 &    86. & -18.8 &  7.56 &  -9.72 \\
J0515-41             &   14.5 &    84. & -17.3 &  7.28 & -10.30 \\
J0516-37             &   18.7 &   140. & -19.8 &  7.51 & -10.05 \\
J0533-36             &   18.4 &   116. & -19.8 &  7.68 &  -9.65 \\
J1002-06             &    9.7 &    64. & -15.9 &  6.84 & -10.05 \\
J1017-03             &   19.4 &   105. & -18.7 &  7.67 &  -9.70 \\
J1046+01             &   12.2 &   118. & -18.3 &  7.31 &  -9.89 \\
J1051-19             &   31.0 &   133. & -20.1 &  7.31 &  -9.94 \\
J1105-00             &    8.6 &   248. & -21.7 &  8.45 &  -9.34 \\
J1107-17             &   11.9 &    68. & -15.7 &  7.64 &  -9.70 \\
J1110+01             &   11.4 &    41. & -15.2 &  7.16 & -10.10 \\
J1127-04             &   10.6 &    43. & -15.1 &  7.11 &  -9.86 \\
J1153-28             &   24.4 &   137. & -20.4 &  7.44 &  -9.82 \\
J1205-14             &   20.4 &   138. & -20.6 &  7.65 &  -9.67 \\
J1217+00             &    8.9 &    32. & -15.0 &  6.60 &  -9.98 \\
J1231-08             &   11.1 &   121. & -19.4 &  7.75 &  -9.26 \\
J1232+00b            &   10.6 &   154. & -20.4 &  7.30 &  -9.78 \\
J1232-07             &   10.5 &   137. & -19.2 &  7.72 &  -9.66 \\
J1235-07             &   10.4 &    95. & -16.6 &  7.38 & -10.01 \\
J1253-12             &    8.6 &    44. & -16.1 &  7.07 &  -9.92 \\
J1255+00             &   15.3 &   116. & -19.1 &  7.51 &  -9.47 \\
J1303-17b            &    7.7 &    56. & -17.1 &  6.83 & -10.16 \\
J1304-10             &   47.4 &   266. & -23.4 &  7.91 &  -9.51 \\
J1329-17             &   22.1 &   257. & -21.8 &  7.60 &  -9.67 \\
J1423+01             &   22.5 &   121. & -19.4 &  7.50 &  -9.62 \\
J1447-17             &   33.5 &   109. & -20.8 &  7.57 &  -9.65 \\
J1500+01             &   22.5 &   191. & -21.2 &  8.21 &  -9.12 \\
J1558-10             &   14.8 &    53. & -16.4 &  7.74 & -10.24 \\
J2009-61             &   11.2 &    62. & -17.1 &  7.28 & -10.04 \\
J2034-31             &   41.9 &   259. & -23.2 &  8.18 &  -9.45 \\
J2044-68             &   45.4 &   230. & -23.1 &  8.02 &  -9.58 \\
J2052-69             &    7.5 &    92. & -18.6 &  7.40 &  -9.72 \\
J2127-60             &   24.9 &   146. & -20.8 &  7.73 &  -9.69 \\
J2129-52             &   12.3 &    81. & -18.3 &  7.44 &  -9.85 \\
J2135-63             &   45.2 &   233. & -23.2 &  8.19 &  -9.32 \\
J2136-54             &   12.0 &   102. & -20.2 &  7.50 &  -9.40 \\
J2214-66             &   24.8 &   184. & -20.5 &  7.63 & -10.39 \\
J2217-42             &   32.1 &    29. & -16.0 &  6.81 & -10.26 \\
J2220-46             &   13.2 &   106. & -19.9 &  7.15 & -10.04 \\
J2234-04             &   14.1 &    53. & -16.6 &  7.12 & -10.12 \\
J2235-26             &   21.1 &   167. & -21.5 &  7.96 &  -9.53 \\
J2239-04             &   13.2 &    33. & -16.0 &  6.90 &  -9.86 \\
J2254-26             &   12.3 &    74. & -16.2 &  7.86 & -10.04 \\
J2257-42             &   13.4 &    59. & -17.0 &  6.65 & -10.13 \\
J2302-40             &   15.3 &    95. & -19.9 &  7.77 &  -9.52 \\
J2326-37             &   10.0 &    37. & -16.5 &  7.37 &  -9.77 \\
J2337-47             &   40.8 &   277. & -22.0 &  7.78 & -10.30 \\
J2349-37             &    9.2 &    45. & -15.3 &  6.62 & -10.18 \\
J2352-52             &    7.7 &    24. & -15.1 &  7.13 &  -9.88 \\

\end{longtable}
\end{center}
Column 1: The HIPASS identification of the galaxy. 
Column 2: Distances in megaparsecs. Column 3: Circular velocity in \kms. Column 4: Dust-corrected optical $R$-band absolute magnitude. Column 5: Effective $R$-band surface brightness density in $L_{\odot}$~kpc$^{-2}$. 
Column 6: Log of \HI-normalised $SFE$.

\twocolumn
\clearpage

\subsection{Comparison with stable disk models }

\begin{figure}
\begin{center}
\includegraphics[scale=.4]{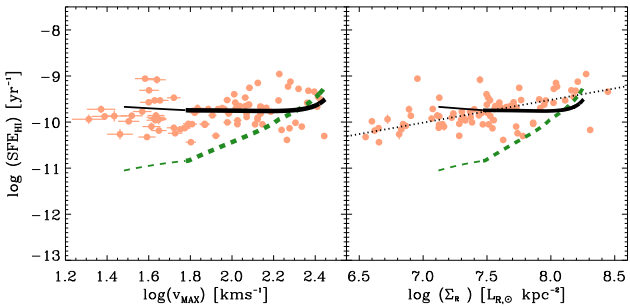}
\end{center}
\caption{The \HI\ star formation efficiency as a function of circular velocity ($v_{\rm{MAX}}$)
 and the effective $R$-band surface brightness ($\Sigma_{\rm{R}}$).
The solid black line represents our model where $R_{\rm{mol}}$ is dictated by the hydrostatic pressure ($P_h$) and the  dashed green line represents our model where the $R_{\rm{mol}}$ depends on the stellar surface density ($\Sigma_{\star}$).  We note that the rotation curves used in our models are calibrated to observations only for \vmax\ covering the range of 60~kms$^{-1}$ to 280~kms$^{-1}$. The thinner line extensions to lower \vmax\ are extrapolations of our disk models' predictions for galaxies with \vmax\ between 30 to 60~kms$^{-1}$.}
\label{sfemodel}
\end{figure}

In Figure~\ref{sfemodel} we compare the observed global $SFE_{\HI}$ for our sample 
of nearby star-forming galaxies to those predicted by our stable disk models (as described in Section 2).
The predicted $SFE_{\HI}$ from our stable disk models are in good agreement to those found for 
our sample of nearby star-forming galaxies.  This is especially the case for our model 
which uses the hydrostatic pressure prescription for $R_{\rm{mol}}$ (equation 10). We 
find a mean log~($SFE_{\HI}$)$=-9.71$ and a standard deviation of 0.02 dex for our stable 
disk model which uses the $P_h$ $R_{\rm{mol}}$ prescription. This mean $SFE_{\HI}$ is evaluated 
individually for each modelled galaxy with \vmax\ between 60 and 280~\kms\ in increments of 0.4~\kms.
While both $R_{\rm{mol}}$ prescriptions yield similar global $SFE_{\HI}$ for galaxies similar in mass 
to the Milky Way (see Figure~\ref{egmodel}d \& Figure~\ref{sfemodel}), the stellar mass density prescription
 (equation 9) for $R_{\rm{mol}}$ predicts that the $SFE_{\HI}$  should correlate  strongly 
with \vmax, which is not observed.

If we extend our models to above our current limit of \vmax$=280$~kms$^{-1}$, we find  sharply
 rising rotation curves, presumably from a strong bulge, resulting in large gas densities 
required to maintain a constant $Q$ and thus high star formation intensity (ie.\ starburst).
Consequently, a sharply rising $SFE_{\rm{HI}}$ with \vmax\ would result for our models
for \vmax$>280$~kms$^{-1}$.   This suggests 
that  either a stable disk cannot be maintained in the central regions
of  high-mass galaxies \citep{zheng13}; or that  high mass galaxies have already 
exhausted their central gas repositories through previous starburst activity. 
We note that for a \vmax$\leq 280$~kms$^{-1}$ limit, the resulting $SFE_{\rm{HI}}$
differs very little from our standard model sequence if we replace our adopted
star formation law with the linear molecular star formation law of \citet{zheng13}.
This is because the required gas densities are typically barely high enough to invoke the 
``starburst'' component of our adopted star formation law, and then only in the central regions of models at
the highest \vmax\ values.  The ``starburst'' 
component becomes more relevant for our low angular momentum model (discussed later
in this section) and our models of $z=2$ galaxies (see Section 5.3).

Within our sample, we do not have any galaxies with \vmax~$>280$~km~s$^{-1}$; they are rare 
in the Local Universe.  As noted in Section 2.1, such galaxies are beyond the range of properties
 used to calibrate the Universal Rotation Curve, hence it is not clear that such an extrapolation
 produces a realistic rotation curve.
Our sample selection excludes galaxies with star-forming companions which may 
further decrease the probability of our sample including galaxies with high \vmax. In addition, higher mass 
systems in the Local Universe are redder and forming less stars \citep[e.g.\ ][]{kauffmann03} 
and hence, not expected to possess large reservoirs of \HI.

While the rotation curve model  used is not calibrated for galaxies with \vmax~$<60$~km~s$^{-1}$,
 the extrapolation of our model (which uses the $P_h$ $R_{\rm{mol}}$ prescription) to lower \vmax\ 
continues to be in good agreement with our observations (see Figure~\ref{sfemodel}).  This suggests 
that the universal rotation curve model adequately accounts for the true rotation curves of dwarf galaxies.  

Our sample of nearby star-forming galaxies show a positive correlation between $SFE_{\HI}$ and the
$R$-band surface brightness, $\Sigma_{R}$. 
A robust linear fit results in a positive slope of 0.50 and an RMS scatter about this fit of 0.23 dex and a Pearson $R$-correlation coefficient of 0.55.  
Relative to our stable disk models, this correlation has a slope which is
less steep than the disk model which uses the $\Sigma_{\star}$ $R_{\rm{mol}}$ prescription (slope of 1.94). 
 While the disk model which uses the $P_h$ $R_{\rm{mol}}$ prescription predicts a better fit to the data, 
the predicted slope is $-0.03$.

\subsection{Variations in the intrinsic surface brightness distribution}

The positive correlation between $\Sigma_{\rm{R}}$ and $SFE_{\HI}$ suggests that a second parameter is at play. Since the primary parameter of our model is \vmax,  a proxy for total mass, an obvious choice for the second parameter is the spin or angular momentum of the disk.  For a given \vmax, a low angular momentum disk will appear more concentrated and hence have a higher surface brightness relative to a high angular momentum disk.  The rotation curve will reach the same \vmax\ but have a steeper rise in the centre.  In the context of our model this results in an increased central star formation rate, but relatively little change in the \HI\ content, and hence a higher $SFE_{\HI}$ for a given \vmax.  Figure~2  shows that the \vmax--surface brightness correlation has a scatter of 0.3 dex in $\Sigma_{\rm{R}}$.  This is equivalent to a scatter of 0.15~dex in scale length, or a factor of 1.4 for pure exponential disks at fixed luminosity. 

\begin{figure}
\begin{center}
\includegraphics[scale=.4]{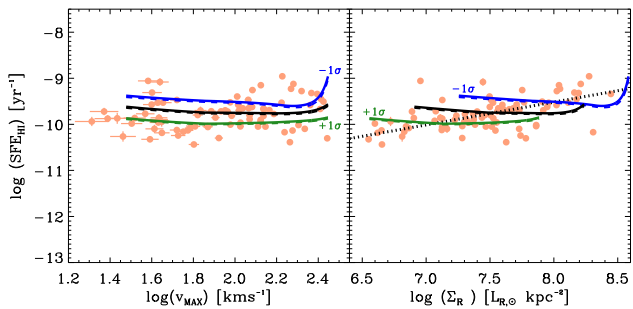}
\end{center}
\caption{The data in these two panels is the same as in Figure~8. Here, the  solid lines represent our model where $R_{\rm{mol}}$ is dictated by the hydrostatic pressure ($P_h$) and the dotted line marks the positive correlation between stellar surface brightness and $SFE_{\HI}$.  The middle solid line (black) correspond to our default model as shown in Figure~9. The top and bottom solid lines correspond to the $-1\sigma$ and $+1\sigma$ variation in the linear scale lengths at fixed stellar luminosities and \vmax, respectively.  {\bf{The dashed lines represent the model with a simple linear molecular star formation law as per \citet{zheng13}.}}}
\label{sfe_grmXpand}
\end{figure}

To test the effects of  angular momentum on our models,  we expanded and contracted the scale length of the disk surface mass distribution, the rotation curve, and the truncation radius while keeping the luminosity and \vmax\ fixed in our models.
 Figure~\ref{sfe_grmXpand} shows another version of Figure~\ref{sfemodel} where the $P_h$ prescription for $R_{\rm{mol}}$
 are plotted in lines of three different thicknesses and colours to represent three variations in our model's surface brightness distribution.  The default surface brightness distribution used in this paper (represented by the black line in Figure~\ref{sfemodel}) is also represented by the black solid line of intermediate thickness in Figure~\ref{sfe_grmXpand}.  The  green and  blue solid lines in Figure~\ref{sfe_grmXpand} represent $\pm$1-$\sigma$  variations in the surface brightness distributions versus \vmax\ relationship.  We note that the scatter in the $R_{\rm{max}}$ versus \vmax\ relationship from which we  derive the constant orbital time is $\approx 0.16$~dex. Therefore, the  scatter in both relationships are consistent with being driven by the scatter in the scale size. In addition, we show in Figure~\ref{sfe_grmXpand} that the differences are not significant between the models that use the star formation law described in Section 2.2 (solid lines) or the linear molecular star formation laws (dashed lines) as per \citet{zheng13}.


As can be seen in the right panel of Figure~\ref{sfe_grmXpand}, by varying the scale length by 0.15~dex and using our $P_h$ prescription for  $R_{\rm{mol}}$ we produce model lines that are shifted diagonally in the $SFE_{\HI}$ versus  $\Sigma_{\rm{R}}$ plane and encompass the envelope of observed data points\footnote{We also examined the effects of varying scale length on the  $\Sigma_{*}$ prescription for $R_{\rm{mol}}$. As in the fiducial model, none of the  $\Sigma_{*}$ prescription models fit the data points, so we do not show the results here so as not to clutter the figures.}
Therefore, we posit that the observed correlation between the stellar surface density and $SFE_{\HI}$ is due to an intrinsic variation in the underlying scale length and hence disk angular momentum of the galaxies.

\section{Discussion}
\subsection{Varying the initial assumptions of our model}
We examine the robustness of the  assumptions and empirical correlations used in our 
models by investigating the resulting effects when each of these assumptions 
 are varied.  We show in this subsection that our results remain unchanged and are
 robust against the uncertainties that exist within the assumptions made and 
empirical correlations used in our model.  Our variation of our initial assumptions
 within known constraints result in models which are still consistent within the
scatter of our observations.  We specifically examine the effects of varying 3
key parameters of our models, namely: a) the value of the assumed constant 
$Q_{\rm{sg}}$; b) the gas velocity dispersion, $\sigma_{\rm{g}}$; and c) the
zero point in the Tully-Fisher relationship.  Table~\ref{varypar} lists the 
observed mean $SFE_{\HI}$ of our sample, the mean $SFE_{\HI}$ of our default model, 
and the mean $SFE_{\HI}$ of the model sequence when each of these parameters are
adjusted.

\begin{table}
\caption{{\bf{Comparison of the observed mean $SFE_{\HI}$ with the mean values of the various models tested in Section 5.1.}}}
\label{varypar}
\begin{center}
\begin{tabular}{lc}
\hline
\hline
Sample  & log (mean $SFE_{\HI}$) [yr$^{-1}$]\\
\hline
Observations &  $-9.65$ ($\pm 0.28$) \\
Default model &  $-9.71$ \\
$Q_{sg}=1.3$ &  $-9.47$ \\
$Q_{sg}=1.9$ &  $-9.83$ \\
$\sigma_g = 8$ \kms\ & $-9.89$\\
$\sigma_g = 20$ \kms\ & $-9.58$\\
1$\sigma$-brighter TFR & $-9.65$\\
1$\sigma$-fainter TFR & $-9.58$\\
\hline
\hline
\end{tabular}
\end{center}
\end{table}

Recent studies such as \citet{zheng13} have measured the radial profiles of the 
Toomre $Q_{\rm{sg}}$ in samples of nearby galaxies and have determined that the 
$Q_{\rm{sg}}$ values within the disks of galaxies are constant with an average 
$Q_{\rm{sg}}$ of 1.6.  However, the standard deviation of the derived $Q_{\rm{sg}}$ 
for the \citet{zheng13} sample is 0.3 excluding the 2 largest outliers in their 
sample. Both outliers in the   \citet{zheng13} sample have $Q_{\rm{sg}} > 2.0$ and 
consist of a dwarf galaxy and a very massive galaxy with \vmax\ of 50~\kms\ and
 300~\kms, beyond the range of the \vmax\ used in the family of rotation curves
 adopted in this paper.    

\begin{figure}
\begin{center}
\includegraphics[scale=.3]{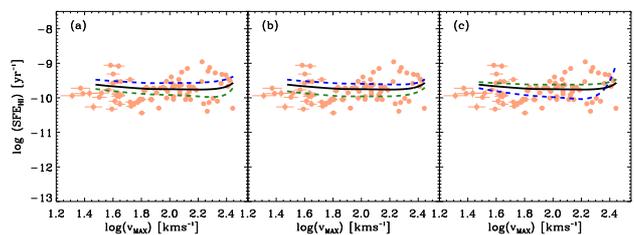}
\end{center}
\caption{The data in these two panels is the same as in Figure~8. Here, the black solid lines
in all 3 panels represent the default model  where $R_{\rm{mol}}$ is dictated by the 
hydrostatic pressure ($P_h$) as in Figure~9.  Panel (a)  shows alternative models which
assume constant $Q_{\rm{sg}}=1.3$ (blue dashed line) and   constant $Q_{\rm{sg}}=1.9$
 (green dashed line). Panel (b)  shows alternative models which assume 
$\sigma_{rm{g}}=14$~\kms\ (blue dashed line) and $\sigma_{rm{g}}=8$~\kms\ (green dashed line). 
Panel (c) shows alternative models which adopt Tully-Fisher relationships which differ
by the $1\sigma$ scatter in Figure~2.  The models which adopt a Tully-Fisher relationship 
that is $1\sigma$ brighter and  $1\sigma$ fainter are represented by the blue dashed 
line and the green dashed line, respectively.
}
\label{sfe_vary}
\end{figure}

Figure~\ref{sfe_vary}a illustrates the effect of varying the assumed constant 
$Q_{\rm{sg}}$ in our models by $\pm 1\sigma$. Increasing to $Q_{\rm sg} = 1.9$ results 
in a lower gas content. The molecular phase and therefore the star formation rate 
is more suppressed than the atomic phase resulting in a lower $SFE_{\HI}$. The converse 
is true if we decrease the stability parameter to $Q_{\rm sg} = 1.3$.

Our adopted constant gas velocity dispersion of 11~\kms\ is based upon 
the mean dispersion measurements from the THINGS \HI\ second-order 
moment maps \citep{leroy08,tamburro09,zheng13}.  The $1\sigma$ scatter
of these measurements is 3~\kms.  We show in Figure~\ref{sfe_vary}b 
that a model which assumes a larger gas velocity dispersion results in 
a higher $SFE_{\HI}$ than one which assumes a lower gas velocity 
dispersion. This is because increasing the velocity dispersion
 allows more gas to be packed in to a disk to get the same $Q_{sg}$, and a 
larger fraction ends up molecular and thus star forming.  The uniformity 
of the resulting $SFE_{\HI}$ as a function of \vmax\ is preserved
 regardless of the assumed velocity dispersion.

Figure~\ref{sfe_vary}c shows the model results for when we vary the 
assumed Tully-Fisher relationship by $1\sigma$ of the observed scatter
in Figure~2.  The model which adopts the Tully-Fisher relationship that is
$1\sigma$ brighter (blue dashed line) shows a general 
suppression in $SFE_{\HI}$ because the brighter Tully-Fisher relationship 
results in higher stellar masses relative to the mean fitted relationship
used.  Our constant $Q_{sg}$ model subsequently balances out the increase
in stellar component with a decrease in the gas component. 
The effect on $SFE_{\HI}$ depends on which phase of the ISM is most affected. 
For \vmax~$\lesssim 220$~\kms\ $SFE_{\HI}$ is less than the default model 
because the molecular phase is more suppressed than the atomic phase of
 the ISM. The situation reverses at larger \vmax, with the \HI\ phase 
more suppressed, resulting in an increased $SFE_{\HI}$.  Models which 
 adopt a fainter Tully-Fisher relationship results in greater gas content 
at all radii, and an overall higher molecular fraction, subsequently yielding 
 greater  $SFE_{\HI}$ than that of our default model at all \vmax.

\subsection{Physical insights from our model}
We see that our marginally stable disk model does a good job of
accounting for the uniform $SFE_{\HI}$ of disk galaxies.  In some sense, the
fact that we can account for the entire sequence of disk galaxies
should not be surprising since it relates other known
properties of the family of star-forming galaxies such as the
Tully-Fisher relation, the surface-brightness luminosity relationship,
an SFL calibrated on disk galaxies, and the
\citet{persic96} URC. The marginal
stability of galaxy disks has also been known for some time
\citep[e.g.\ ][]{kennicutt89,leroy08,meurer13}.  However, this is the first time
it has been invoked as the primary physical basis for understanding
the sequence of star forming galaxies.

We find that the $P_h$ prescription for $R_{\rm mol}$ works best for
explaining the uniform $SFE_{\HI}$.  Previously \citet{leroy08} found
that an empirical scaling of $R_{\rm mol}$ with stellar mass density
$\Sigma_\star$ worked marginally better than the more physically based
$P_h$ prescription.  However, that study was limited to the optically
bright part of galaxies and did not include  the outer disks where \HI\
becomes an important contributor to the disk mass density, and where
weak star formation can often  be seen, especially in the UV
\citep{gildepaz07,thilker07}.  Even where the star formation intensity in
such an extended gaseous disk is weak compared to the central star
formation, the contribution to the total SFR can be significant.  A
case in point is the BCD galaxy NGC~2915, where the outer disk star formation
intensity is $\sim 0.5$\%\ that in the centre of
this galaxy yet comprises $\sim$11\%\ of the
total SFR \citep{bruzzese15,werk10}.

Our assumed stellar mass distribution is a pure exponential disk, i.e.\ without a bulge. The results indicate this model is adequate.  It is probable that the \HI\ selection of our sample ensures that we do not have many bulge-dominated or early-type galaxies. The sample used to define our rotation curve model  was selected to have Hubble types Sb to Sd, thus avoiding the more bulge dominated Sa types \citep{mathewson92}. In the context of our models, there are two other reasons why bulge dominated galaxies are unlikely to contribute greatly to the URC. Firstly, the high central densities of bulge dominated disks require very high $\Sigma_g$ to produce $Q = 1.6$.  Consequently, the star formation in this regime will be at the starburst portion of our adopted SFL, resulting in a short gas consumption timescale, hence making these galaxies rare. Secondly, those galaxies that do manage to accrete enough gas into their centres to match our model would need to have done so quickly, i.e.\ through an interaction or merger. This should lead to large scale velocity field asymmetries, and galaxies with asymmetric velocity profiles were excluded from the sub-sample of galaxies used by \citet{persic95,persic96} to compile the URC.

The model presented here is an elaboration of our previous work on the
implications of disks maintaining constant stability.  In
\citet{meurer13} we showed that the gas dominated outer disks with a
constant $Q$ in potentials with a flat rotation curve can match the
observed radial $\Sigma_{\rm HI}$ distributions and gas to total mass
ratio. In \citet{zheng13} we
showed that by assuming a constant \Qsg, we could crudely reproduce
radial profiles of \HI, \Htwo, and $\Sigma_{\rm SFR}$ out to the
traditional optical extents of galaxies given their rotation curves and
stellar mass profiles. Since star formation is more centrally
concentrated than \HI, here we are required to consider the full radial
properties of galaxies.

We note that our models have some similarities to the semi-analytic
models of \citet{lagos11}.  They also used a hydrostatic pressure
relationship to apportion the ISM in galaxies into molecular and atomic
phases and a molecular star formation law to determine the SFR.  One
key difference is that they assume that $\Sigma_g$ has an exponential
radial profile.  They allow the disk mass and scale
length to evolve to accommodate the build-up of gas content and angular
momentum through accretion. They are then able to reproduce a large
variety of properties of disk galaxies. However, $SFE_{\HI}$ is not one of
the properties that they studied. Their models are more maturely developed than ours
which address neither the source of the gas nor their evolution. The
strength of our model is the physical basis for the structure of the
disk. Hence, it should be straightforward to adapt our model for evolutionary
effects.




\subsection{Application to distant newly-formed galaxies}
As early as redshift $z \sim 2$, disk-dominated systems seem to be well
established, albeit with  higher velocity dispersions relative to galaxies in the 
Local Universe. Recent studies of star-forming galaxies at $z \sim 2$
have found while the morphologies appear clumpier than low-$z$ counterparts
and with higher velocity dispersions,
 a significant fraction of galaxies are disk-dominated and can be described
 by a relatively constant $Q$ \citep{genzel08,cresci09,foerster09,genzel14}.
Following the results of \citet{foerster09}, we find an average 
$\sigma_g = 60\, {\rm km\, s^{-1}}$, a median H$\alpha$ half light radius of
4.6~kpc, a median SFR=135\,\msun\,yr$^{-1}$,  a median effective star formation
 intensity (SFR per unit projected area within the half light radius)
 of $0.74\, \msun\, {\rm kpc^{-2}\,  yr^{-1}}$, a median \vmax=235~\kms and
 a median dynamical mass of $1.4\times 10^{11}$~M$_{\odot}$.
 How do these properties compare to what we would expect
for young massive disk galaxies?

\begin{figure}
\begin{center}
\includegraphics[scale=.6]{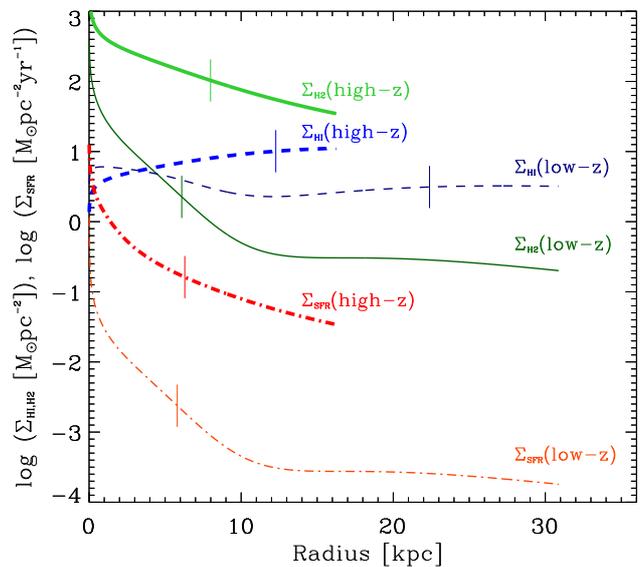}
\end{center}
\caption{A comparison of the gas density and star formation intensity radial profiles of our Milky Way
analogue models (\vmax=220~\kms) at low- and high-redshifts. The \HI\ and \Htwo\ densities are represented
by the dashed and solid lines, respectively, while the star formation intensity is represented by the
dot-dash line. The thick and thin versions of each linestyle represent the high and low redshift models.
The effective radius for each profile is marked by a thin vertical solid line. }
\label{sfe_highz}
\end{figure}

To this end, we consider the properties of a Milky
Way analogue, having $\vmax = 220$~km~s$^{-1}$, at the early stages of its
formation. Our adopted  model for this \vmax\ for a galaxy in the Local 
Universe, presented in
\S~\ref{s:demon}, has a total baryon mass, taken to be the sum of the stellar and
gas disk mass, $M_{\rm baryon} = 8.3\times 10^{10}\, \msun$, of which
18\%\ is gas. The total gas mass is thus $1.5\times 10^{10}\, \msun$ where 
the atomic component account for  84\%\ of the total gas.  For our young galaxy
 model we place this entire baryon mass in a pure gas disk having $Q = 1.6$ and 
$\sigma_g = 60\,{\rm km\, s^{-1}}$.  We adopt the same rotation curve model as before,
and now use the familiar single fluid $Q_g$ model to determine the
$\Sigma_g$ distribution. As before, we use the hydrostatic pressure
prescription to apportion the gas into the atomic and molecular phases and
the molecular SFL (Section 2.1) to derive the the resultant radial
star formation distribution.  The radial profiles are truncated when the
enclosed gas mass equals the above $M_{\rm baryon}$ (see Figure~\ref{sfe_highz}). 
We find that the gas in this high-$z$ model would be predominantly molecular, 
while 10\%\ is atomic ($M_{\rm HI}= 8.0\times 10^9\, \msun$)---much less than 
that found for the low-$z$ model.

From these radial profiles we derive an integrated SFR of 94~\msun\,yr$^{-1}$, a 
star formation half-light radius to be 6.3~kpc, and an effective star formation 
intensity of $0.38\, \msun\, {\rm kpc^{-2}\,  yr^{-1}}$. 
Our modelled integrated SFR is very close to the median observed
SFR of disk dominated galaxies at $z\sim 2$, while 
the effective star formation intensity is within the observed range of these galaxies
 \citep[e.g.\ ][]{johnston15,foerster09}. 
On the other hand, the size of the modelled galaxy is slightly larger than observed. 
Table~\ref{comparez} summarises the main properties found for 
our models.
 
We note that if only half of the galaxy had assembled at these high redshifts (i.e. the outer
50\%\ of the gas mass and the resultant star formation were removed),
then the galaxy model would extend out to 8.9~kpc, have a SFR of 60~\msun\ yr$^{-1}$, 
a SFR effective radius of 4.1~kpc, and an effective star
formation intensity of $0.56\, \msun\, {\rm kpc^{-2}\, yr^{-1}}$. In
this case all quantities are within the range observed for the $z \sim 2$ disk galaxies. 
Therefore,  our model proto-Milky Way would have star forming properties more similar to what is observed in high-redshift
young disk galaxies, if  half to all of its baryons are assembled
in a disk having $Q \sim 1.6$.

\begin{table}
\caption{Comparison of model properties for a Milky-Way-like disk galaxy (\vmax=220~\kms) at low- and high redshifts.}
\label{comparez}
\scriptsize{
\begin{center}
\begin{tabular}{llcc}
\hline
\hline
Property   & $z=0$ & $z=2$\\
\hline
\HI\ effective radius  &  22.4~kpc & 12.3~kpc\\
\Htwo\ effective radius & 6.1~kpc & 8.0~kpc\\
$SFR$  effective radius & 5.8~kpc & 6.3~kpc\\
$SFR$ &  1.8 ~$M_{\odot}$~yr$^{-1}$ & 94~$M_{\odot}$~yr$^{-1}$\\
Total gas depletion time, $t$ & 6.3 Gyr  & 0.84 Gyr \\
log~$SFE_{\HI}$ & $-9.72$~year$^{-1}$ & $-7.91$~year$^{-1}$\\
$\Sigma_{\rm{SFR(R_e)}}$ & 0.008~$M_{\odot}$ kpc$^{-2}$ yr$^{-1}$   & 0.377~$M_{\odot}$ kpc$^{-2}$ yr$^{-1}$\\
\hline
\hline
\end{tabular}
\end{center}}
\end{table}

\section{Conclusions}

The sequence of star forming galaxies defined by our sample, spans five orders of magnitude
 in stellar mass and is unimodal. We find that the constant $SFE_{\HI}$  along this sequence can be effectively described by galaxy disks with a constant marginal stability.
  The primary parameter giving the structure of a galaxy in this model is the 
rotational amplitude \vmax, a proxy for the total galaxy mass.
  Combined with previously-known scaling relations (such as the Tully-Fisher relationship,
and the surface brightness--\vmax\ relationship), we derive the distribution of
 the stars and dark matter. The gas disk structure, the relative distribution of the
gas phases and subsequently the star formation is a consequence of this framework.
While the internal structure of galaxies varies greatly along the sequence, the integrated
$SFE_{\rm{HI}}$ of galaxies is constant at the truncation radii of disks.

A weak correlation between the $SFE_{\rm{HI}}$ and $\Sigma_{R}$ indicates that angular momentum
 is a second parameter affecting the distribution of the ISM and stars.  Low specific 
angular momentum galaxies have shorter scale lengths and a more compact mass distribution.  
In the context of our models, this results in higher mass density gaseous disks and 
consequently, higher surface brightnesses.  Similarly, lower specific angular momentum 
galaxies have less dense disks and less intense star formation.  Our conclusions are
 similar to  those of \citet{popping15} who found the overall $SFE$ 
(SFR normalised by total gas) of galaxies to be primarily governed by the 
absolute amount of cold gas; with a secondary dependence upon the disk scale lengths.

Previous studies \citep[e.g.\ ][]{leroy08,wong13} have emphasised  
 the observed correlation between $R_{\rm{mol}}$ (and $\Sigma_{H2}$) with $\Sigma_{*}$ ---
suggesting that the local supply of molecular gas is regulated by the stellar disk of
the galaxy. However the observed correlations linking $\Sigma_{*}$ to $\Sigma_{\rm{SFR}}$ is 
not unexpected \citep[e.g.\ ][]{ryder94,dopita94} because the hydrostatic pressure and the
 stellar surface density are strongly correlated within the optical disks of massive galaxies 
 \citep[e.g.\ ][]{blitz06,leroy08}. The fact that the 
 observations are well-modelled with the hydrostatic  pressure model suggests that it
 provides a better overall prescription for setting the \Htwo/\HI\ ratio and hence, 
the \HI-based $SFE_{\HI}$. This is especially the case for dwarf galaxies where the 
gaseous disk (as opposed to the stellar disk) is more relevant for setting the hydrostatic
 pressure.

In this paper, we have also shown that our simple constant-$Q$ model is able to reproduce
the observed star formation properties of a proto-Milky Way at higher redshifts when 
the Universe was at its peak star formation period.  Since our model is static and only
 deals with the disk component, we have not been able to address the effects from evolution or 
other common galactic structures such as bulges and bars.  However, the success of our model
at reproducing the observed star formation properties of low- and high-redshift galaxies 
 suggest that star formation in galaxies is largely due to the formation and stabilisation 
 of their disks.








\vspace{1cm}

\chapter{\bf{Acknowledgments.}}
Support for the work presented here was also obtained
through a NASA GALEX Guest Investigator grant, GALEX GI04-0105-0009
 and GALEX archival grant NNX09AF85G. Early work on the data presented
 here was also supported by NASA LTSA grant NAG5-13083 to G.R. Meurer.
We thank the anonymous referee for their support of this paper and 
for improving the manuscript. 
This research has made use of the NASA/IPAC Extragalactic Database (NED),
which is operated by the Jet Propulsion Laboratory, California Institute of Technology,
under contract with the National Aeronautics and Space Administration.

\bibliographystyle{mnras}
\bibliography{mn-jour,paperef}

\end{document}